\documentclass[%
 reprint,
 amsmath,
 amssymb,
 aps,
 prl,
 twocolumn
]{revtex4-1}

\usepackage{graphicx}
\usepackage{dcolumn}
\usepackage{bm}
\usepackage{appendix}

\usepackage{graphicx}
\usepackage{dcolumn}
\usepackage{bm}
\usepackage{algorithm}
\usepackage{algpseudocode}

\usepackage{capt-of}

\usepackage{xr}
\begin{document}

\newcommand{\boldeta}{\eta} 
\newcommand{\boldlambda}{{\boldsymbol{\lambda}}}
\newcommand{\boldx}{{{x}}}
\newcommand{\trajx}{X}
\newcommand{\boldc}{c}
\newcommand{\dt}{{\,\mathrm{d}t}}
\newcommand{\ds}{{\,\mathrm{d}s}}
\newcommand{\dx}{{\,\mathrm{d}x}}
\newcommand{\dee}{{\mathrm{d}}}
\newcommand{\gammadot}{{\dot{\gamma}}}

\title{
Beyond Linear Response: Equivalence between Thermodynamic Geometry and Optimal Transport
}
\author{Adrianne Zhong$^{1,2,}$}
\email{adrizhong@berkeley.edu}
\author{Michael R. DeWeese$^{1,2,3}$}
\affiliation{%
${^1}$Department of Physics, University of California, Berkeley, Berkeley, CA, 94720 \\
${^2}$Redwood Center For Theoretical Neuroscience, University of California, Berkeley, Berkeley, CA, 94720 \\
${^3}$Department of Neuroscience, University of California, Berkeley, Berkeley, CA, 94720
}%

\date{\today}

\begin{abstract}
A fundamental result of thermodynamic geometry is that the optimal, minimal-work protocol that drives a nonequilibrium system between two thermodynamic states in the slow-driving limit is given by a geodesic of the friction tensor, a Riemannian metric defined on control space. For overdamped dynamics in arbitrary dimensions, we demonstrate that thermodynamic geometry is equivalent to $L^2$ optimal transport geometry defined on the space of equilibrium distributions corresponding to the control parameters. We show that obtaining optimal protocols past the slow-driving or linear response regime is computationally tractable as the sum of a friction tensor geodesic and a counterdiabatic term related to the Fisher information metric. These geodesic-counterdiabatic optimal protocols are exact for parameteric harmonic potentials, reproduce the surprising non-monotonic behavior recently discovered in linearly-biased double well optimal protocols, and explain the ubiquitous discontinuous jumps observed at the beginning and end times. 
\end{abstract}

\maketitle

\emph{Introduction.}---A consequence of the Second Law of Thermodynamics is that finite-time processes require work to be irretrievably lost as dissipation. Recent studies in stochastic thermodynamics have aimed to characterize minimal-work protocols, which have applications for nanoscopic engineering \cite{diana2013finite, zulkowski2014optimal, proesmans2020finite, proesmans2020optimal, whitelam2023train, schmiedl2007efficiency, martinez2017colloidal, martin2018extracting, abiuso2020optimal, brandner2020thermodynamic, frim2022optimal, frim2022geometric} and for understanding biophysical systems \cite{geiger2010optimum, dellago2013computing, sivak2016thermodynamic, lucero2019optimal,  blaber2022efficient, davis2024active}. In this Letter we unify disparate geometric approaches and arrive at a novel framework for obtaining and better understanding thermodynamically optimal protocols.

The problem statement is: Given a configuration space $x \in \mathbb{R}^d$, inverse temperature $\beta$, and potential energy function $U_\lambda(x)$ parameterized by $\lambda \in \mathcal{M}$, what is the optimal protocol $\lambda^*(t)$ connecting the parameter values $\lambda_i$ and $\lambda_f$ in a finite time $\tau$ that minimizes the work 
\begin{equation}
  W[\lambda(t)] = \int_0^\tau \frac{\dee \lambda^\mu}{\dt} \bigg\langle \frac{\partial U_{\lambda}}{\partial \lambda^\mu } \bigg\rangle \dt \label{eq:work-definition} \, ?
\end{equation}
Here $\mathcal{M} \subseteq \mathbb{R}^n$ is an orientable $m$-dimensional manifold, locally resembling $\mathbb{R}^m$ everywhere with $m \leq n$. We use Greek indices to denote local coordinates of $\lambda \in \mathcal{M}$, and the Einstein summation convention (i.e., repeated Greek indices within a term are implicitly summed). The ensemble average $\langle \cdot \rangle$ is over trajectories $X(t)|_{t \in [0, \tau]}$ that start in equilibrium with $\lambda_i$ and evolve via some specified Hamiltonian or Langevin dynamics under $U_{\lambda(t)}|_{t \in [0, \tau]}$. 

Schmiedl and Seifert \cite{schmiedl2007optimal} showed that optimal protocols minimizing Eq.~\eqref{eq:work-definition} have intriguing \textit{discontinuous jumps} at the beginning and end times, which have proven to be ubiquitous ~\cite{schmiedl2007optimal,jump-inclusion-in-integral, underdamped-regularity,then2008computing, bonancca2018minimal, naze2022optimal, blaber2021steps, zhong2022limited, whitelam2023train, rolandi2023optimal, engel2023optimal, esposito2010finite}. Furthermore, optimal protocols can even be \textit{non-monotonic} in time \cite{zhong2022limited, whitelam2023train}. 

Sivak and Crooks demonstrated through linear response \cite{zwanzig2001nonequilibrium} that in the slow-driving limit ($\tau \gg \tau_\mathrm{R}$, an appropriate relaxation time-scale), optimal protocols are \textit{geodesics} of a symmetric positive-definite~\cite{trivial-exception} friction tensor defined in terms of equilibrium time-correlation functions. Treating the friction tensor as a Riemannian metric induces a geometric structure on the space of control parameters, known as ``Thermodynamic geometry.'' This approach is computationally tractable, as the friction tensor can be obtained through measurement, and geodesics on $\mathcal{M}$ can be determined by solving an ordinary differential equation. Geodesic protocols have been studied for a variety of systems including the Ising model \cite{rotskoff2015optimal, gingrich2016near, rotskoff2017geometric, louwerse2022multidimensional}, barrier crossing \cite{sivak2016thermodynamic, lucero2019optimal, blaber2022efficient}, bit-erasure \cite{zulkowski2014optimal, scandi2022minimally}, and nanoscopic heat engines (after allowing temperature to be controlled) \cite{abiuso2020optimal, brandner2020thermodynamic, frim2022optimal, frim2022geometric, chennakesavalu2023unified}, but unfortunately their performance can degrade past the slow-driving regime \cite{zhong2022limited}. 

Alternatively, when the ensemble of trajectories is additionally constrained to end in equilibrium with $\lambda_f$, finding the work-minimizing protocol for overdamped dynamics is equivalent to the Benamou-Brenier formulation of the $L^2$ optimal transport problem~\cite{aurell2011optimal, ito2024geometric}---finding the dynamical mapping between the two distributions that has minimal integrated squared distance \cite{benamou2000computational}---which itself yields a Riemannian-geometric structure \cite{otto2001geometry, kniefacz2017otto, villani2021topics}. The Benamou-Brenier solution is a time-dependent distribution and time-dependent velocity field that solves a continuity equation, which in this Letter we explicitly identify as a desired probability density evolution and the additional counterdiabatic forcing needed to effectuate its faster-than-quasistatic time evolution (as studied in so-called Engineered Swift Equilibriation \cite{martinez2016engineered, baldassarri2020engineered, frim2021engineered}, Counterdiabatic Driving \cite{iram2021controlling}, and Shortcuts to Adiabaticity \cite{torrontegui2013shortcuts, jarzynski2013generating, campbell2017trade, guery2019shortcuts, plata2021taming, guery2023driving, patra2021semiclassical, li2017shortcuts, li2022geodesic}). Remarkably, optimal protocols obtained in this manner are exact for arbitrary protocol durations $\tau$~\cite{aurell2011optimal}. Unfortunately, this approach involves solving coupled PDEs in configuration space ($\mathbb{R}^d$), which is typically infeasible for dimension $d \gtrsim 5$. Furthermore, the control space $\mathcal{M}$ must be sufficiently expressive in order to implement the optimal transport solution, which is often overly-restrictive \cite{klinger2024universal}. 

For overdamped dynamics in arbitrary dimension, \cite{chennakesavalu2023unified} showed that the friction tensor may be obtained via a perturbative expansion of the Benamou-Brenier objective function. Here we derive an even stronger result, that thermodynamic geometry is in fact \textit{equivalent} to optimal transport geometry, in the sense that the friction tensor and the Benamou-Brenier problem restricted to equilibrium distributions parameterized by $\lambda$ have identical geodesics and geodesic distances. Surprisingly, we find that a counterdiabatic component may be calculated using the Fisher information metric from information geometry \cite{fisher1922mathematical, rao1992information, amari2016information}. We demonstrate that protocols obtained by adding this counterdiabatic term to thermodynamic geometry geodesics are analytically exact for parametric harmonic oscillators, reproduce recently discovered non-monotonic behavior in certain optimal protocols \cite{zhong2022limited}, and satisfyingly explain the origin of jumps at beginning and end times.

\begin{figure} 
\centering
\includegraphics[width=8cm
]{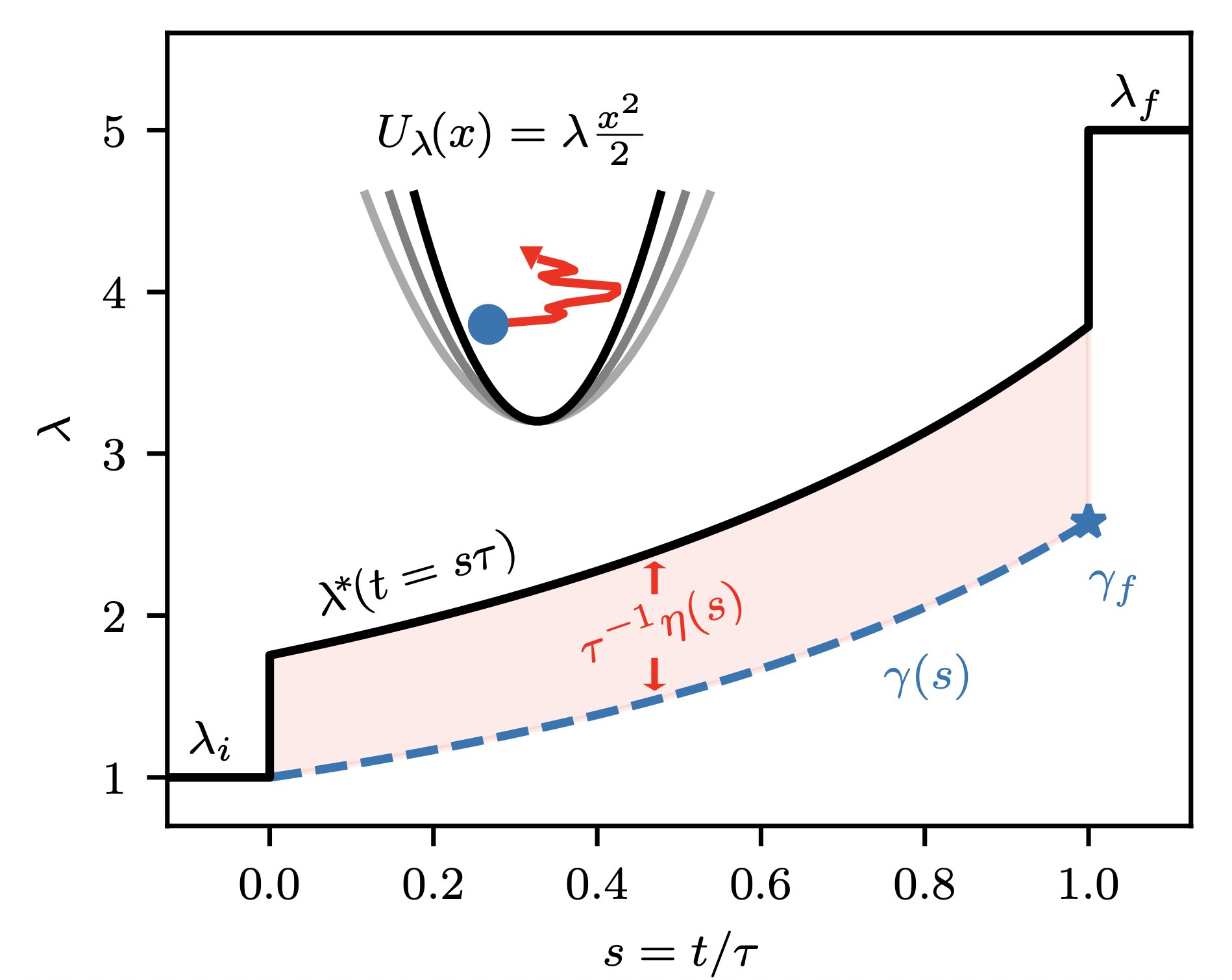}
\caption{Alg.~1. reproduces the exact optimal protocol (black) for the variable-stiffness harmonic oscillator (depicted here, $\lambda_i = 1$, $\lambda_f = 5$, and $\tau = 0.5$) found in \cite{schmiedl2007optimal}, as the sum of geodesic (dashed blue) and counterdiabatic (shaded red) components. The geodesic $\gamma(s)$ connects $\gamma(0) = \lambda_i$ to $\gamma(1) = \gamma_f$ (blue star) solving Eq.~\eqref{eq:gamma-f}. 
} 
\label{fig:fig1}
\end{figure}

\emph{Preliminaries.}---For each $\lambda \in \mathcal{M}$ there is a corresponding equilibrium distribution 
\begin{equation}
    \rho^\mathrm{eq}_\lambda(x) = \exp \{ - \beta[U_\lambda(x) - F(\lambda)] \}, \label{eq:equilibrium-dist}
\end{equation}
where $F(\lambda) = -\beta^{-1} \ln \int \exp[-\beta U_\lambda(x')] \dx'$ is the equilibrium free energy of the potential energy $U_\lambda(\cdot)$. For ease of notation we will denote $\rho_i^\mathrm{eq} = \rho^\mathrm{eq}_{\lambda_i}$, $\rho_f^\mathrm{eq} = \rho^\mathrm{eq}_{\lambda_f}$, and $\Delta F = F(\lambda_f) - F(\lambda_i)$. 

We consider overdamped Langevin equations, such that
trajectories $X(t) \in \mathbb{R}^d$ follow the stochastic ODE
\begin{gather}
  \dee X(t) = - \nabla U_{\lambda(t)}(X(t)) \dt  + \sqrt{2 \beta^{-1}} \dee B(t), \label{eq:langevin-eq}
\end{gather}
where $B(t) \in \mathbb{R}^d$ is an instantiation of standard Brownian motion \cite{oksendal2013stochastic}. Here we will consider only isothermal protocols, so WLOG we set $\beta = 1$. 

The probability density $\rho(x, t)$ corresponding to Eq.~\eqref{eq:langevin-eq} undergoes a time-evolution expressible either as a Fokker-Planck equation or a continuity equation of a gradient field 
\begin{equation}
    \frac{\partial \rho }{\partial t} = \mathcal{L}_{\lambda(t)} \rho \quad\mathrm{or}\quad \frac{\partial \rho }{\partial t} = \nabla \cdot (\rho \nabla \phi )  \label{eq:fokker-planck-eq}, 
\end{equation}
where $\mathcal{L}_\lambda$ is the Fokker-Planck operator~\cite{risken1996fokker}
\begin{equation}
    \mathcal{L}_\lambda \rho = \nabla^2 \rho + \nabla \cdot (\rho \, \nabla U_\lambda), \label{eq:fokker-planck-operator}
\end{equation}
while $\phi$ is a scalar field that depends on both $\rho$ and $\lambda$
\begin{equation}
  \phi(x, t) = \ln \rho(x, t) + U_{\lambda(t)}(x, t). \label{eq:effective-velocity}
\end{equation}
The adjoint operator $\mathcal{L}_\lambda^\dagger$ acts on a scalar field $\psi(x)$ via~\cite{BK-form}
\begin{align}
    [\mathcal{L}^\dagger_\lambda \psi] (x) &= \rho_\lambda^\mathrm{eq}(x)^{-1} \nabla \cdot [ \rho_\lambda^\mathrm{eq} (x)  \nabla \psi(x) ].  \label{eq:BK-1} 
\end{align}
Finally, $f_\mu(x) := - \partial U_\lambda(x) / \partial \lambda^\mu$ is the conjugate force to $\lambda^\mu$. The excess conjugate force is then 
\begin{equation}
    \delta f_\mu(x) = - \bigg[ \frac{\partial U_\lambda(x)}{\partial \lambda^\mu} - \bigg\langle \frac{\partial U_\lambda}{\partial \lambda^\mu} \bigg\rangle_\lambda^\mathrm{eq} \bigg]  = \frac{\partial \ln \rho_\lambda^\mathrm{eq}(x)}{\partial \lambda^\mu}. \label{eq:excess-conj-force}
\end{equation}

\emph{Thermodynamic geometry.}---In the slow-driving limit, the excess work, defined as the work (Eq.~\eqref{eq:work-definition}) minus the equilibrium free energy difference $W_\mathrm{ex} = W - \Delta F$, is~\cite{sivak2012thermodynamic}  
\begin{equation}
  W_\mathrm{ex}[\lambda(t)] \approx \int_0^\tau \frac{\dee \lambda ^\mu}{\dt}  \frac{\dee \lambda^\nu }{\dt}  g_{\mu \nu} (\lambda(t)) \dt, \label{eq:SC-cost}
\end{equation}
where
\begin{equation}
  g_{\mu \nu}(\lambda) = \int_0^\infty \big\langle \delta f_\mu(X(t')) \, \delta f_\nu(X(0))  \big\rangle^\mathrm{eq}_\lambda \dt' \label{eq:friction-SC}
\end{equation}
is the symmetric positive-definite ~\cite{trivial-exception} friction tensor. Here, $\langle \cdot \rangle^\mathrm{eq}_\lambda$ denotes an equilibrium average (i.e., $X(0) \sim \rho_\lambda^\mathrm{eq}$, and trajectories undergo Langevin dynamics (Eq.~\eqref{eq:langevin-eq}) with constant $\lambda$).

Remarkably, the friction tensor induces a Riemannian geometry on control space $(\mathcal{M}, g)$ known as ``Thermodynamic geometry,'' with squared thermodynamic length between $\lambda_A, \lambda_B \in \mathcal{M}$ given by minimizing the path action
\begin{gather}
    \mathcal{T}^2 (\lambda_A, \lambda_B) = \min_{\lambda(s) |_{s \in [0, 1]}} \bigg\{ \int_0^1  \frac{\dee \lambda ^\mu}{\ds}  \frac{\dee \lambda^\nu }{\ds}  g_{\mu \nu} (\lambda(s)) \ds \ \bigg| \nonumber \\
    \mathrm{satisfying \ } \lambda(0) = \lambda_A, \lambda(1) = \lambda_B \bigg\} . \label{eq:thermodynamic-distance-SC}
\end{gather}
In the slow-driving limit, optimal protocols $\lambda^*(t)$ connecting $\lambda_i$ and $\lambda_f$ in time $\tau$ are time-rescaled versions of geodesics of Eq.~\eqref{eq:thermodynamic-distance-SC}, and the optimal excess work scales inversely with protocol time $W_\mathrm{ex}^* \approx \mathcal{T}^2 (\lambda_i, \lambda_f) / \tau$ \cite{sivak2012thermodynamic, zulkowski2012geometry}. 

While this geometric framework is both mathematically elegant and computationally tractable, geodesic protocols are fundamentally approximate; their performance often degrades for sufficiently small protocol times, in some cases performing even worse than a linear interpolation protocol \cite{zhong2022limited}.

\emph{Optimal transport geometry.}---Optimal transport is traditionally formulated as finding the transport map sending a distribution $\rho_A$ to another $\rho_B$ that minimizes an integrated ($L^2$) squared distance. This minimal integrated squared distance defines 
the squared $L^2$-Wasserstein metric distance between probability distributions, which was shown in \cite{benamou2000computational} to also be the minimum of a path action 
\begin{widetext}
   \begin{equation}
   \mathcal{W}_2^2[\rho_A, \rho_B] = \min_{\rho_s, \phi_s |_{s \in [0,1]}} \bigg\{ \int_0^{1} \int \rho_s(x) | \nabla \phi_s (x)|^2 \dx \ds \ \bigg| \ \mathrm{satisfying \ } \frac{\partial \rho_s}{\partial s} = \nabla \cdot (\rho_s \nabla \phi_s), \, \rho_0 = \rho_A, \, \rho_1 = \rho_B \bigg\}. \label{eq:BB-cost-function}
   \end{equation}
\end{widetext}
Here $\rho_s(\cdot)|_{s \in [0,1]}$ is a trajectory of configuration space probability densities $\mathcal{P}(\mathbb{R}^d)$ \cite{finite-second-moment}, and $\phi_s(\cdot)|_{s \in [0, 1]}$ is a trajectory of scalar fields that yield gradient velocity fields $v_s = -\nabla \phi_s$ satisfying the continuity equation $\partial_s \rho_s = - \nabla \cdot (\rho_s v_s)$ \cite{OT-meticulous}. This so-called Benamou-Brenier formulation of optimal transport (Eq.~\eqref{eq:BB-cost-function}) reveals a Riemmanian structure on the space of probability distributions known as Otto calculus \cite{otto2001geometry, kniefacz2017otto, villani2021topics}: on this manifold of probability distributions $M := \mathcal{P}(\mathbb{R}^d)$, a ``point'' is a probability distribution $\rho \in M$, a ``tangent space vector'' is a gradient velocity field identifiable (up to a constant offset) by a scalar field $\phi \in T_\rho(M)$, and geodesics are the argmin of Eq.~\eqref{eq:BB-cost-function}.

For overdamped dynamics, the work-minimizing protocol satisfying boundary conditions $\rho(\cdot, 0) = \rho_i^\mathrm{eq}$ and $\rho(\cdot, \tau) = \rho_f^\mathrm{eq}$ is a time-scaled solution of Eq.~\eqref{eq:BB-cost-function} for $\rho_A = \rho_i^\mathrm{eq}, \rho_B = \rho_f^\mathrm{eq}$, assuming sufficiently expressive control \cite{aurell2011optimal}. (See the SM for a concise derivation.) From the continuity equation form of $\partial \rho(\cdot, t) / \partial t$ (Eqs.~\eqref{eq:fokker-planck-eq} and~\eqref{eq:effective-velocity}), the optimal protocol $\lambda^*(t)$ for finite time $\tau$ can be expressed in terms of $\rho_s^*$ and $\phi_s^*$ that solve Eq.~\eqref{eq:BB-cost-function}, as satisfying (up to a constant offset)
\begin{equation}
  U_{\lambda^*(t)}(x) = -\ln \rho_{t/\tau}^*(x) + \tau^{-1} \phi_{t/\tau}^*(x) .\label{eq:OT-solution}
\end{equation}
The first term corresponds to the Benamou-Brenier geodesic $\rho_{s}^*$, and the second one with $\phi_{s}^*$ is a counterdiabatic term that drives the probability distribution solving Eq.~\eqref{eq:fokker-planck-eq} to match the geodesic $\rho(\cdot, t) = \rho^*_{t / \tau}$ (see \cite{counterdiabatic-driving-correspondence}).

Remarkably, this solution is \textit{exact} for any finite $\tau$, and it provides a geometric interpretation for these work-minimizing protocols as optimal transport geodesics connecting $\rho_i^\mathrm{eq}$ to $\rho_f^\mathrm{eq}$. Through the time-scaling $t = \tau s$, it follows that $W_\mathrm{ex}^* = \mathcal{W}_2^2[\rho_i^\mathrm{eq}, \rho_f^\mathrm{eq}] / \tau$ is a \textit{tight} lower bound for excess dissipation in this additionally-constrained setting \cite{dechant2019thermodynamic, nakazato2021geometrical}. However, there are two important caveats to this approach: first, solving Eq.~\eqref{eq:BB-cost-function} involves PDEs on configuration space, which generally for dimension $d \gtrsim 5$ is computationally intractable (although, see \cite{beck2021solving, han2018solving, chen2021solving, albergo2023stochastic} for sophisticated modern machine learning methods, as well as \cite{klinger2024universal}). Second, the control parameters must be sufficiently expressive in the sense that for all $t \in (0, \tau)$ there \text{has to be} a $\lambda \in \mathcal{M}$ that satisfies Eq.~\eqref{eq:OT-solution}. Worse yet, there might not be \textit{any} admissible protocols that can satisfy the terminal constraint $\rho(\cdot, \tau) = \rho_f^\mathrm{eq}$~\cite{klinger2024universal}. 

Without the terminal condition, this problem is no longer over-constrained. The optimal excess work can be expressed as a minimum over $\rho_f = \rho(\cdot, \tau)$ (see the SM), 
\begin{equation}
    W_\mathrm{ex}^* = \min_{\rho_f} \ \mathcal{W}_2^2[\rho_i^\mathrm{eq}, \rho_f]  /  \tau + D_\mathrm{KL}(\rho_f \, | \, \rho^\mathrm{eq}_f) , \label{eq:free-endpoint}
\end{equation}
where the additional KL-divergence cost
\begin{equation}
    D_\mathrm{KL}(\rho_A \, | \, \rho_B) := \int \rho_A (x)  \ln  \frac{\rho_A(x)}{\rho_B(x)}  \dx , \label{eq:KL-divergence}
\end{equation}
is the dissipation from the equilibration $\rho_f \rightarrow \rho^\mathrm{eq}_f$ that occurs for $t > \tau$ (see \cite{JKO}). Optimal protocols $\lambda^*(t)$ are also obtained via Eqs.~\eqref{eq:BB-cost-function} and~\eqref{eq:OT-solution}, but now with $\rho_A = \rho_i^\mathrm{eq}$ and $\rho_B = \rho_f^*$ that minimizes Eq.~\eqref{eq:free-endpoint}. Without the restrictive terminal constraint, protocols that approximate Eq.~\eqref{eq:OT-solution} are allowed (in the case of limited expressivity), and may be near-optimal in performance \cite{gingrich2016near, engel2023optimal}.

\emph{Demonstrating equivalence of geometries.}---We start by expressing Eq.~\eqref{eq:friction-SC} with the time-propagator (e.g., see Ch 4.2 of \cite{risken1996fokker}):
\begin{align}
 g_{\mu \nu}(\lambda)\ &= \int_0^\infty \int \rho_\lambda^\mathrm{eq}(x) \, \delta f_\mu(x) \, e^{\mathcal{L}_\lambda^\dagger t'} \, \delta f_\nu(x)  \dx \dt' \nonumber \\
 &=  -\int \rho_\lambda^\mathrm{eq}(x)  \, \delta f_\mu(x) \, \big\{\mathcal{L}_\lambda^\dagger \big\}^{-1}\, [\delta f_\nu ]  (x) \dx.  \label{eq:friction-tensor-3}
\end{align}
The second line comes from taking the time-integral, where the inverse operator $\big\{\mathcal{L}_\lambda^\dagger \big\}^{-1}$ is defined in terms of a properly constructed Green's function (Eq.~(40) in \cite{wadia2022solution}). This expression is the lowest order tensor found in a perturbative expansion of the Fokker-Planck equation~\cite{wadia2022solution}.

By formally defining $\phi_\mu = \big\{\mathcal{L}_\lambda^\dagger \big\}^{-1} \delta f_\mu$ as (up to a constant offset) the scalar field solving $\mathcal{L}_\lambda^\dagger \phi_\mu = \delta f_\mu$, it is straightfoward to show with Eqs.~\eqref{eq:BK-1} and~\eqref{eq:excess-conj-force} that, for any protocol $\lambda(s)|_{s \in [0, 1]}$,
\begin{equation}
    \frac{\partial \rho_{\lambda(s)}^\mathrm{eq} }{\partial s} = \nabla \cdot ( \rho_{\lambda(s)}^\mathrm{eq} \nabla \phi_s), \  \ \mathrm{where} \ \ \phi_s(x) = \frac{\dee \lambda^\mu}{\ds} \phi_\mu(x).  \label{eq:friction-phi}
\end{equation}
Applying $\delta f_\mu = \mathcal{L}^\dagger_\lambda \phi_\mu$ and Eq.~\eqref{eq:BK-1} to Eq.~\eqref{eq:friction-tensor-3} shows that the thermodynamic distance (Eq.~\eqref{eq:thermodynamic-distance-SC}) may be expressed as
\begin{widetext}
    \begin{equation}
    \mathcal{T}^2(\lambda_A, \lambda_B) = \min_{\lambda(s), \phi_s|_{s \in [0, 1]}} \bigg\{ \int_0^{1} \int \rho_{\lambda(s)}^\mathrm{eq}(x) |\nabla \phi_s(x)|^2 \dx \ds \ \bigg| \ \mathrm{satisfying \ } \frac{\partial \rho_{\lambda(s)}^\mathrm{eq}}{\partial s} = \nabla \cdot (\rho_{\lambda(s)}^\mathrm{eq} \nabla \phi_s), \, \lambda(0) = \lambda_A, \lambda(1) = \lambda_B \bigg\}.
    \label{eq:squared-thermodynamic-length-2}
    \end{equation} 
\end{widetext}
This is our first major result: this expression is equivalent to the squared $L^2$-Wasserstein distance (Eq.~\eqref{eq:BB-cost-function}) with the constraint that $\rho_s|_{s \in [0, 1]}$ is a trajectory of equilibrium distributions $\rho^\mathrm{eq}_{\lambda(s)}|_{s \in [0, 1]}$. In other words, thermodynamic geometry induced by the friction tensor (Eq.~\eqref{eq:friction-SC}) on $\mathcal{M}$ is \textit{equivalent} to optimal transport geometry restricted to the equilibrium distributions $\mathcal{P}^\mathrm{eq}_\mathcal{M}(\mathbb{R}^d)$ corresponding to $\mathcal{M}$ \cite{ball}, and thus share the same geodesics and geodesic distances.

Up till now, thermodynamic geometry has prescribed optimal protocols as friction tensor geodesics joining $\lambda_i$ and $\lambda_f$, which are approximate for finite $\tau$. Optimal transport solutions require solving PDEs, but yield exact optimal protocols containing both geodesic and counterdiabatic components (Eq.~\eqref{eq:OT-solution}). Our unification of geometries suggests that thermodynamic geometry protocols may be made exact by including a counterdiabatic term.

\emph{Geodesic-counterdiabatic optimal protocols.}---From here we consider the control-affine parameterization
\begin{equation}
  U_\lambda(x) = U_\mathrm{fixed}(x) + U_\mathrm{offset}(\lambda) + \lambda^\mu U_\mu (x), \label{eq:affine-U}
\end{equation}
and control space $\lambda \in \mathcal{M} = \mathbb{R}^m$. 
It follows from the equivalence of thermodynamic and optimal transport geometries that the optimal protocol should have the form
\begin{equation}
    \lambda^*(t) = \gamma(t / \tau) + \tau^{-1} \eta(t / \tau), \label{eq:counterdiabatic-protocol}
\end{equation}
namely the sum of a geodesic term and a counterdiabatic term that correspond to the two terms in  Eq.~\eqref{eq:OT-solution}, where $\rho_s^*|_{s\in[0,1]}$ and $\phi^*_s|_{s\in[0,1]}$ solve Eq.~\eqref{eq:BB-cost-function} with $\rho_A = \rho_i^\mathrm{eq}$ and $\rho_B = \rho_f^*$ from Eq.~\eqref{eq:free-endpoint}. Here, $\gamma(s)$ will be a geodesic of $g(\lambda)$ joining $\gamma(0) = \lambda_i$ to $\gamma(1) = \gamma_f$, where
\begin{equation}
  \gamma_f = \arg \min_{\lambda} \ \mathcal{T}^2(\lambda_i, \lambda) / \tau + D_\mathrm{KL}(\rho_\lambda^\mathrm{eq} | \rho_f^\mathrm{eq} ).
  \label{eq:gamma-f}
\end{equation}
We show in Appendix A that the counterdiabatic term is
\begin{equation} \label{eq:eta-solution}
    \eta(s) = h^{-1}(\gamma(s)) \, g(\gamma(s))  \bigg[ \frac{\dee \gamma(s)}{\ds} \bigg], 
\end{equation}
where, intriguingly, $h$ is the Fisher information metric \cite{fisher-trad-expression}
\begin{equation}
  h_{\mu \nu}(\lambda) = \int \rho_\lambda^\mathrm{eq}(x) \, 
  \delta f_\mu(x) \, \delta f_\nu(x) \dx, \label{eq:fisher-information-metric}
\end{equation}
which also induces a Riemannian geometry on the space of parametric equilibrium probability distributions $(\mathcal{M}, h)$ known as ``Information geometry'' \cite{fisher1922mathematical, rao1992information, amari2016information}. Eq.~\eqref{eq:eta-solution} is exact in cases of sufficient expressivity (i.e., when Eq.~\eqref{eq:OT-solution} can be satisfied); otherwise, $\eta(s)$ is the full solution projected onto $\mathcal{M}$. 

This is our second major result: the equivalence between thermodynamic and optimal transport geometries implies that optimal protocols beyond linear response require counterdiabatic forcing, and can be obtained via:
\begin{algorithm}[H] \label{alg:algorithm}
  \caption{Geodesic-counterdiabatic opt. protocols. \\ 
  \textbf{Input:} $\lambda_i, \lambda_f$, protocol time $\tau$, metrics $g_{\mu \nu}(\lambda)$, $h_{\mu \nu}(\lambda)$ (Eqs.~\eqref{eq:friction-SC}, \eqref{eq:fisher-information-metric}), KL divergence $D_\mathrm{KL}(\cdot \, |\,  \rho_f^\mathrm{eq})$ (Eq.~\eqref{eq:KL-divergence}).}
   \begin{algorithmic}[1]
   \State Solve geodesic $\gamma(s)|_{s\in[0,1]}$ connecting $\gamma(0) = \lambda_i$ and $\gamma(1) = \gamma_f$ (obtained from Eq.~\eqref{eq:gamma-f}) under $g_{\mu \nu}$.
   \State Calculate counterdiabatic term $\eta(s) = h^{-1} g \, [\dee \gamma / \dee s]$
   \State Return optimal protocol $\lambda^*(t) = \gamma(t / \tau) + \tau^{-1} \eta(t / \tau)$
   \end{algorithmic}
\end{algorithm}
We emphasize that this procedure does not require solving any configuration-space PDEs. Moreover, in the limit $\tau \rightarrow \infty$, the counterdiabatic component in Eq.~\eqref{eq:counterdiabatic-protocol} vanishes and Eq.~\eqref{eq:gamma-f} is solved by $\gamma_f = \lambda_f$, and thus geodesic protocols from thermodynamic geometry are reproduced.

\emph{Examples.}---We show in Appendix B that Alg.~1. reproduces exact optimal protocols solved in \cite{schmiedl2007optimal} for controlling a parameteric harmonic potential. Fig. 1. illustrates an optimal protocol for $U_\lambda(x) = \lambda x^2 / 2$. Notice that at $t = 0$ the counterdiabatic term is suddenly turned on, while at $t = \tau$ the geodesic ends at $\gamma_f \neq \lambda_f$ and the counterdiabatic term is suddenly turned off. Seen in this light, the discontinuous jumps in optimal protocols $\lambda^*(t)$ arise from the sudden turning on and off of counterdiabatic forcing, and the discontinuity of the geodesic at $t = \tau$. We note that starting in equilibrium at $t = 0$ and suddenly ending control at $t = \tau$ are both unnatural in biological settings; these generic discontinuous jumps can be seen as artifacts of the imposed boundary conditions. 

\begin{figure}
\includegraphics[width=8.5cm
]{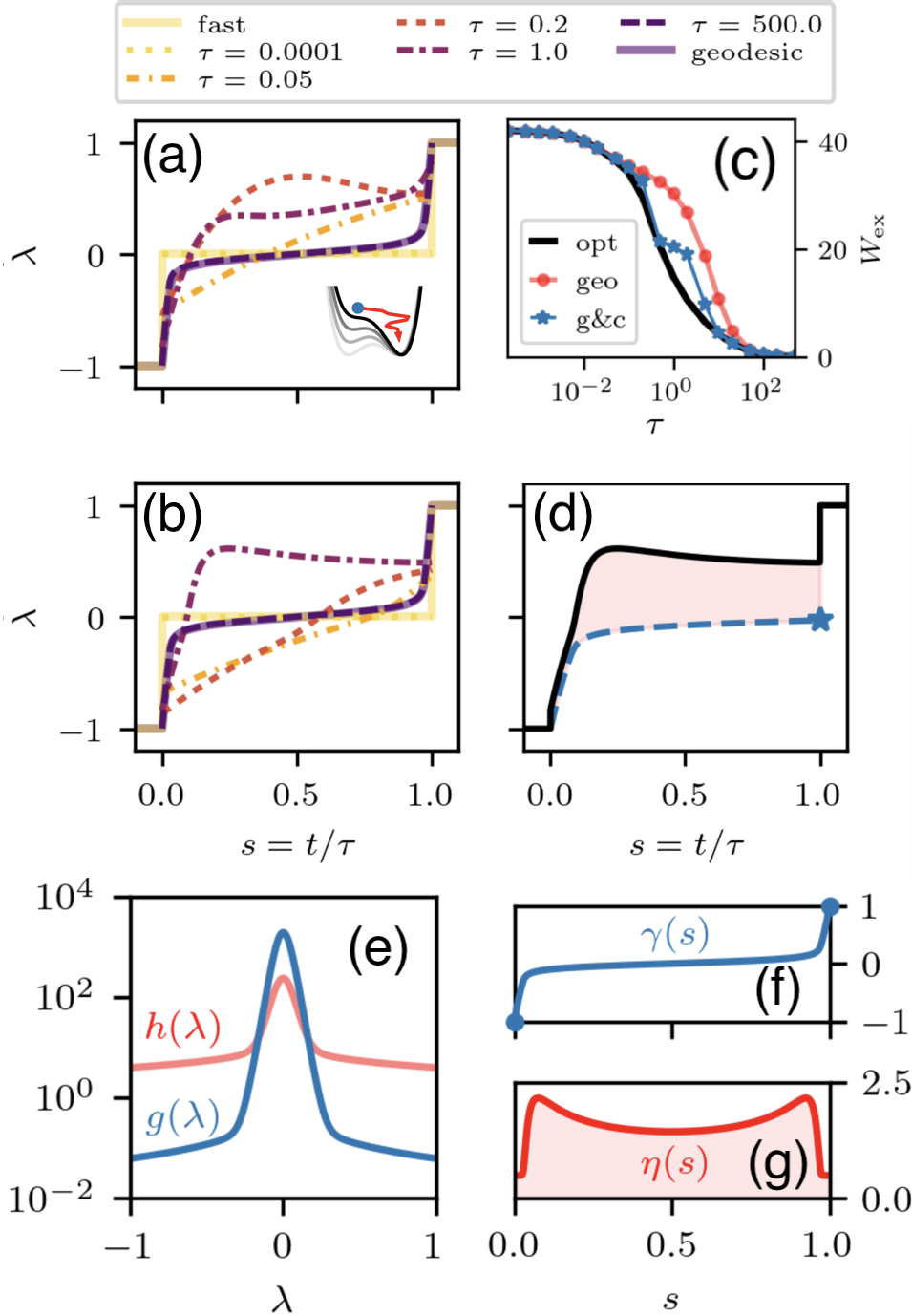}
\caption{(a) Exact optimal protocols obtained in \cite{zhong2022limited} from solving PDEs, for the linearly-biased double well (Eq.~\eqref{eq:double-well-potential}; $E_0 = 16$) for different protocol durations $\tau$ including the fast protocol $\tau \rightarrow 0$ \cite{blaber2021steps} (solid yellow) and the friction tensor geodesic protocol (dark purple). (b) Geodesic-counterdiabatic protocols numerically obtained from Alg.~1. (c) Geodesic-counterdiabatic protocols (blue stars) outperform the geodesic protocol (red circles) for all $\tau$; cf. performance of exact optimal protocols (black). We examine the reduction in performance at $\tau \sim 2$ in the SM. (d) The $\tau = 1$ protocol numerically obtained via Alg.~1 (here $\gamma_f = 0.0291$), same coloring as Fig.~1. (e) The friction and Fisher information tensors yield (f) the geodesic $\gamma(s)$ (here $\lambda_A = -1$, $\lambda_B = 1$) and (g) the non-monotonic
counterdiabatic forcing $\eta(s)$.}
\label{fig:fig2}
\end{figure}

Surprisingly, \emph{non-monotonic} optimal protocols have been found for the linearly-biased double well ~\cite{zhong2022limited} 
\begin{equation}
  U_\lambda(x) = E_0[(x^2 - 1)^2 / 4 - \lambda x],   \label{eq:double-well-potential}
\end{equation}
for certain values of $E_0$ and $\tau$ (e.g., $\tau = 0.2$ in Fig.~2(a) for $E_0 = 16$). Fig.~2(b) illustrates protocols numerically obtained from Alg.~1 (details given in the SM). Due to the limited expressivity of the controls, these protocols are not identical to the exact optimal protocols obtained by solving PDEs~\cite{zhong2022limited} (Fig.~2(a)). However, they reproduce significant properties (e.g., discontinuous jumps and non-monotonicity, becoming exact in $\tau \rightarrow 0$ and $\tau \rightarrow \infty$), and lead to improved performance over geodesic protocols (Fig.~2(c)). Fig.~2(d) illustrates the non-monotonic $\tau = 1$ protocol as a sum of geodesic and counterdiabatic terms. The tensors $g$ and $h$ (Fig.~2(e)) yield necessarily monotonic geodesics (Fig.~2(f)), and non-monotonic counterdiabatic forcing (Fig.~2(g)) that leads to non-monotonic optimal protocols. 

\emph{Discussion.}---We have demonstrated the equivalence between overdamped thermodynamic geometry on $\mathcal{M}$---previously seen as an approximate framework---and $L^2$ optimal transport geometry on equilibrium distributions $\mathcal{P}^\mathrm{eq}_\mathcal{M}(\mathbb{R}^d) \subset \mathcal{P}(\mathbb{R}^d)$. The resulting geodesic-counterdiabatic optimal protocols from Alg.~1 are exact for parameteric harmonic traps, and explain both the ubiquitous discontinuous jumps and the non-monotonic behavior observed in optimal protocols. 

We note that \cite{li2022geodesic} presents a geodesic-counterdiabatic PDEs approach for underdamped dynamics. Additionally, underdamped optimal control has recently been related to a modified optimal transport problem \cite{sabbagh2023wasserstein, sanders2024optimal}. We expect that the metric tensor in \cite{li2022geodesic}, the friction tensor (Eq.~\eqref{eq:friction-SC}) for underdamped dynamics \cite{zulkowski2012geometry}, and the optimal transport specified in \cite{sabbagh2023wasserstein, sanders2024optimal} may also be geometrically unified through methods similar to ours. 

An interesting future direction will be to apply our findings to heat engines \cite{abiuso2020optimal, brandner2020thermodynamic, frim2022optimal, frim2022geometric, chennakesavalu2023unified} and active matter systems \cite{zulkowski2013optimal, davis2024active}, which have been studied with approximate geodesic protocols. We hope that the insight that minimal-work protocols require both geodesic and counterdiabatic components will prove to be useful in understanding the cyclic and fundamentally non-equilibrium processes that occur in biological systems. 

\begin{acknowledgments}
\emph{Acknowledgements.}---This work greatly benefited from conversations with Adam Frim.  A.Z. was supported by the Department of Defense (DoD) through the National Defense Science \& Engineering Graduate (NDSEG) Fellowship Program.
MRD thanks Steve Strong and Fenrir LLC for supporting this project. This work was supported in part by the U.S. Army Research Laboratory and the U.S. Army Research Office under Contract No. W911NF-20-1-0151.

\emph{Appendix A: Counterdiabatic driving expression.---}In this Appendix we derive our expression for the counterdiabatic term (Eq.~\eqref{eq:eta-solution}). Per definition, the counterdiabatic term $\phi_s(x) = \eta^\mu(s) U_\mu(x)$ is constructed to solve the continuity equation 
\begin{equation}
  \frac{\partial \rho_{\gamma(s)}^\mathrm{eq} }{\partial s}(x) = \nabla \cdot ( \rho_{\gamma(s)}^\mathrm{eq}(x) \nabla \phi_s(x)) .\label{eq:counterdiabatic-PDE}
\end{equation}
We can divide by $\rho^\mathrm{eq}_\lambda(x)$ and plug in $\partial_s \ln \rho_{\gamma(s)}^\mathrm{eq}  = [ \dee \gamma(s) / \ds] \, \delta f_\mu(x) $  to obtain
\begin{equation}
    \frac{\dee \gamma^\mu}{\ds} \delta f_\mu(x) = - \eta^\nu(s) [\mathcal{L}_{\gamma(s)}^\dagger \, \delta f_\nu ](x), \label{eq:counterdiabatic-1}
\end{equation}
where we have used the fact that $U_\nu(x) = -\delta f_\nu + \mathrm{const}$, and that the adjoint operator (Eq.~\eqref{eq:BK-1}) satisfies $\mathcal{L}^\dagger_\lambda [\psi + c] = \mathcal{L}^\dagger_\lambda [\psi]$ for any scalar field $\psi(x)$ and constant $c \in \mathbb{R}$.

Due to the limited expressivity of available controls in a given problem, it might not be possible to satisfy Eq.~\eqref{eq:counterdiabatic-1}. However, this potential insolubility is resolved by applying a projection operator to both sides
\begin{widetext}
\begin{equation}
    \frac{\dee \gamma^\mu}{\ds}  \int \bigg[ - \rho^\mathrm{eq}_{\gamma(s)}(x) \, \delta f_\alpha (x) \, \big\{ \mathcal{L}_{\gamma(s)}^\dagger \big\}^{-1} \bigg] \delta f_\mu(x) \dx = - \eta^\nu(s) \int \bigg[ - \rho^\mathrm{eq}_{\gamma(s)}(x) \, \delta f_\alpha (x) \, \big\{ \mathcal{L}_{\gamma(s)}^\dagger \big\}^{-1} \bigg] \big[\mathcal{L}_{\gamma(s)}^\dagger \, \delta f_\nu \big](x) \dx, 
\end{equation}
\end{widetext}
which yields 
\begin{equation}
    g_{\alpha \mu}(\gamma(s)) \bigg[ \frac{\dee \gamma^\mu}{\ds} \bigg] = h_{\alpha \nu} (\gamma(s)) \, \eta^\nu(s), \label{eq:counterdiabatic-2}
\end{equation}
using the friction tensor and Fisher information metric expressions Eq.~\eqref{eq:friction-tensor-3} and Eq.~\eqref{eq:fisher-information-metric}. Optimal transport geometry measures ``horizontal'' displacement while information geometry measures ``vertical'' displacement \cite{khan2022optimal}, so $g$ and $h$ can be seen to give the conversion between the left and right hand sides of Eq.~\eqref{eq:counterdiabatic-PDE} \cite{choice-in-projection}.

Because the Fisher information metric is symmetric and positive-definite (assuming \cite{trivial-exception}), we can apply its inverse to Eq.~\eqref{eq:counterdiabatic-2}, thus reproducing Eq.~\eqref{eq:eta-solution}. 

\emph{Appendix B: Analytic reproduction of harmonic oscillator optimal protocols.---}In this Appendix we show that Alg. 1 exactly reproduces the optimal protocols for harmonic potentials first found in \cite{schmiedl2007optimal}. 

We first consider a variable-center harmonic potential in one dimension $U_\lambda(x) = (x - \lambda)^2/2$, both the friction and Fisher information tensors are spatially constant $g(\lambda) = f(\lambda) = 1$ due to translation symmetry. For any $\lambda_f$, the KL-divergence given by $D_\mathrm{KL}(\rho_\lambda^\mathrm{eq} | \rho_f^\mathrm{eq} ) = \int \rho_\lambda^\mathrm{eq}(x) \ln \,  [ \rho_\lambda^\mathrm{eq}(x) / \rho_f^\mathrm{eq}(x) ] \dx = (\lambda - \lambda_f)^2 / 2$, while the squared thermodynamic length is given by $\mathcal{T}^2(\lambda_i, \lambda) = (\lambda - \lambda_i)^2$. Without loss of generality we fix $\lambda_i = 0$. Following Alg. 1, we obtain $\gamma_f = (1 + 2 / \tau)^{-1} \lambda_f$ with geodesic $\gamma(s) = s \, \gamma_f$, and $\eta(s) = \gamma_f$. Ultimately, this yields the optimal protocol
\begin{equation}
    \lambda^*(t) = \underbrace{[ \lambda_f / (2 + \tau)] \, t}_\text{geodesic} \hspace{5pt}  + \underbrace{ 1 / (2 + \tau)}_\text{counterdiabatic}\hspace{-5pt},  \quad\mathrm{for}\quad t \in (0, \tau), 
\end{equation}
which yields the original analytic solution reported by Schmiedl and Seifert $\lambda^*(t) = \lambda_f(1 + t) / (2 + \tau)$, Eq.~(9) in \cite{schmiedl2007optimal} (note that they use $t$ to denote protocol duration and $\tau$ to denote time, which is swapped with respect to our notation). However, we now have a refined interpretation of this optimal protocol: as consisting of a geodesic component that connects $\lambda_i$ to $\gamma_f \neq \lambda_f$, and a counterdiabatic component necessary to achieve the geodesic trajectory for finite protocol durations $\tau$.

We now solve for the optimal protocol for a variable-stiffness harmonic trap $U_\lambda(x) = \lambda x^2 / 2$. We will defer solving for $\gamma_f$ until after obtaining the analytic form for $\lambda^*(t)$. The friction tensor for this potential has previously been shown to be $g(\lambda) = 1/4 \lambda^3$ \cite{sivak2012thermodynamic}, and the Fisher information metric can be calculated to be $h(\lambda) = \big\langle x^4 / 4 \big\rangle_\lambda^\mathrm{eq} - \big( \langle x^2 / 2 \rangle_\lambda^\mathrm{eq} \big)^2  = 1/2 \lambda^2$. As shown in \cite{sivak2012thermodynamic}, by switching to standard deviation coordinates $\sigma = \lambda^{-1/2}$, the friction tensor is constant $\tilde{g}(\sigma) = 1$, and thus geodesics are $\sigma(s) = (1 - s) \, \sigma_A + s \, \sigma_B$ with thermodynamic length $\mathcal{\tilde{T}}^2(\sigma_A, \sigma_B) = (\sigma_B - \sigma_A)^2$, where $\sigma_A = \lambda_i^{-1/2}$ and $\sigma_B = \gamma_f^{-1/2}$. This yields the geodesic
\begin{equation}
    \gamma(s) = [(1 - s) \sigma_A + s \, \sigma_B ]^{-2},
\end{equation}
with the corresponding counterdiabatic term given by
\begin{equation}
  \eta(s) = (\sigma_A - \sigma_B) / [(1 - s) \sigma_A + s \, \sigma_B ].
\end{equation}
Plugging these into $\lambda^*(t) = \gamma(t / \tau) + \tau^{-1} \eta (t / \tau)$ leads to the interpretable optimal protocol expression
\begin{equation}
  \lambda^*(t) = \underbrace{\big[ 1 / \sigma^*(t) \big]^{2} }_\text{geodesic} + \underbrace{\tau^{-1} \big[ \Delta \sigma / \sigma^*(t) \big] }_\text{counterdiabatic}, \label{eq:solution-HO-stiffening}
\end{equation}
where $\sigma^*(t) = \sigma_i + (t / \tau) \Delta \sigma $ is the linear interpolation between the endpoints $\sigma_i = \lambda_i^{-1/2}$ and $\sigma_i + \Delta \sigma = \gamma_f^{-1/2}$. 

Finally we solve for $\gamma_f$ via
\begin{equation}
    \gamma_f = \arg \min_{\lambda} \ \mathcal{T}^2(\lambda_i, \lambda) / \tau + D_\mathrm{KL}(\rho_\lambda^\mathrm{eq} | \rho_f^\mathrm{eq} ), \label{eq:gamma_f_def}
\end{equation}
using $\mathcal{T}^2(\lambda_i, \lambda) = (\lambda_i^{-1/2} - \lambda^{-1/2})^2$ (recall that $\lambda^{-1/2} = \sigma$ is the standard deviation) and $D_\mathrm{KL}(\rho_\lambda^\mathrm{eq} | \rho_f^\mathrm{eq} ) = (1/2)[(\lambda_f/\lambda - 1) + \ln (\lambda / \lambda_f)]$ \cite{7440}. This yields
\begin{equation}
    \gamma_f =  \big( \sqrt{1 + 2 \lambda_i \tau + \lambda_i \lambda_f \tau^2 } - 1 \big)^2 / \lambda_i \tau^2,
\end{equation}
or when expressed in $\sigma$ coordinates
\begin{equation}
    \Delta \sigma = \sigma_i  \bigg( \frac{ 1 + \lambda_f \tau - \sqrt{1 + \lambda_k \tau + \lambda_i \lambda_f \tau^2 }}{ 2 + \lambda_f \tau } \bigg) . \label{eq:delta-sigma}
\end{equation}
Substituting this expression into Eq.~\eqref{eq:solution-HO-stiffening} reproduces the exact optimal protocol, Eqs.~(18) and~(19) in \cite{schmiedl2007optimal}. (Note that they use $t$ to denote protocol duration and $\tau$ to denote time, which is swapped with respect to our notation.)

Fig.~1 illustrates the obtained exact optimal protocol for this problem (Eqs.~\eqref{eq:solution-HO-stiffening} and~\eqref{eq:delta-sigma}) as a sum of geodesic and counterdiabatic components. This clarifies the origin of the jumps in optimal protocols: at $t = 0$, the counterdiabatic component is suddenly turned on, and at $t = \tau$ it is abruptly turned off.

\end{acknowledgments}

\bibliography{main}

\begin{thebibliography}{89}
\bibitem [{jum()}]{jump-inclusion-in-integral}%
  \BibitemOpen
  \href@noop {} {}\bibinfo {note} {The discontinuities in the protocol are also taken into account in the integral Eq.~\eqref{eq:work-definition}}\BibitemShut {NoStop}%
\bibitem [{tri()}]{trivial-exception}%
  \BibitemOpen
  \href@noop {} {}\bibinfo {note} {We assume the parameterization of $U_\lambda$ is such that $\rho_{\lambda}^\mathrm{eq}$ are different for different $\lambda$; otherwise, the tensor could be semi-positive definite.}\BibitemShut {Stop}%
\bibitem [{bal()}]{ball}%
  \BibitemOpen
  \href@noop {} {}\bibinfo {note} {This is entirely analogous to how both the unit ball $B^2 = \{ (x, y, z) \, | \, x^2 + y^2 + z^2 \leq 1 \}$ and unit sphere $S^2 = \{ (x, y, z) \, | \, x^2 + y^2 + z^2 = 1 \}$ can be seen as submanifolds of Euclidean space $\mathbb{R}^3$; while geodesics connecting two points in the unit ball are equivalent to straight-line Euclidean geodesics, on the unit sphere they differ.}\BibitemShut {Stop}%
\bibitem [{fis()}]{fisher-trad-expression}%
  \BibitemOpen
  \href@noop {} {}\bibinfo {note} {Traditionally the Fisher information metric is denoted as $h_{\mu \nu}(\lambda) = \int \rho_\lambda^\mathrm{eq} \, (\partial \ln \rho_\lambda^\mathrm{eq} / \partial \lambda^\mu) \, (\partial \ln \rho_\lambda^\mathrm{eq} / \partial \lambda^\nu) \dx$ (which can be seen as equivalent via Eq.~\eqref{eq:excess-conj-force}).}\BibitemShut {Stop}%
\bibitem [{cho()}]{choice-in-projection}%
  \BibitemOpen
  \href@noop {} {}\bibinfo {note} {One may choose other projection operators to act on Eq.~\eqref{eq:counterdiabatic-1}, but then non-conventional tensors would have to be defined. Furthermore, the particular interpretation for Eq.~\eqref{eq:counterdiabatic-2}, that the geometries associated with $g$ and $h$ convert between horizontal and vertical displacements, will be lost.}\BibitemShut {Stop}%
\end{thebibliography}
\begin{thebibliography}{8}
\bibitem[{fir()}]{first-law-of-thermodynamics}
\bibinfo{note}{Note, this identity is equivalently the First Law of Thermodynamics $\dee E / \dt = \dee Q / \dt + \dee W / \dt$.}
\end{thebibliography}

\end{document}


\newcommand{\boldeta}{\eta} 
\newcommand{\boldlambda}{{\boldsymbol{\lambda}}}
\newcommand{\boldx}{{{x}}}
\newcommand{\trajx}{X}
\newcommand{\boldc}{c}
\newcommand{\dt}{{\,\mathrm{d}t}}
\newcommand{\ds}{{\,\mathrm{d}s}}
\newcommand{\dx}{{\,\mathrm{d}x}}
\newcommand{\dee}{{\mathrm{d}}}
\newcommand{\gammadot}{{\dot{\gamma}}}

\newcommand{\boldrho}{{\boldsymbol{\rho}}}
\newcommand{\boldU}{{\boldsymbol{U}}}
\newcommand{\boldpi}{{\boldsymbol{\pi}}}
\newcommand{\boldpsi}{{\boldsymbol{\psi}}}
\newcommand{\curlyL}{{\cal{L}}}
\newcommand{\curlyW}{{\cal{W}}}

\newcommand{\tee}{{\mathrm{t}}}
\newcommand{\tf}{{\mathrm{t_f}}}
\newcommand{\A}{{\mathrm{A}}}
\newcommand{\B}{{\mathrm{B}}}
\newcommand{\C}{{\mathrm{C}}}
\newcommand{\D}{{\mathrm{D}}}
\newcommand{\U}{{\mathrm{U}}}
\newcommand{\W}{{\mathrm{W}}}
\newcommand{\bolddx}{{\Delta \boldsymbol{x}}}

\preprint{APS/123-QED}

\title{Supplementary Information for ``Beyond Linear Response: Equivalence between Thermodynamic Geometry and Optimal Transport''
}
\author{Adrianne Zhong$^{1,2,}$}
\email{adrizhong@berkeley.edu}
\author{Michael R. DeWeese$^{1,2,3}$}
\affiliation{%
${^1}$Department of Physics, University of California, Berkeley, Berkeley, CA, 94720 \\
${^2}$Redwood Center For Theoretical Neuroscience, University of California, Berkeley, Berkeley, CA, 94720 \\
${^3}$Department of Neuroscience, University of California, Berkeley, Berkeley, CA, 94720
}%

\date{\today}
\maketitle

\section{Optimal transport formulation of optimal, work-minimizing protocols}

We present a concise derivation the optimal transport formulation of minimum-work protocols. For simplicity here we use the notation $\rho_t = \rho(\cdot, t)$ and $U_t = U_{\lambda(t)}(\cdot)$. For now, we consider the initial conditions $\rho_0 = \rho_i^\mathrm{eq}$ and $U_0 = U_i$, and the terminal condition $U_\tau = U_f$, \textit{without} imposing $\rho_\tau = \rho_f^\mathrm{eq}$. 

Recall that the Fokker-Planck equation giving the time-evolution for $\rho_t$ may be written as the continuity equation 
\begin{equation}
  \frac{\partial \rho_t}{\partial t} = \nabla \cdot (\rho_t \nabla \phi_t), \quad\quad\mathrm{where}\quad\quad \phi_t :=  U_t + \ln \rho_t. \label{eq:FP-continuity}
\end{equation}

The ensemble work rate is defined as 
\begin{equation}
    \frac{\dee W}{\dt} = \int \rho_t(x) \, \frac{\dee U_t(x)}{\dt} \dx.
\end{equation}
By applying the identity $\frac{\dee}{\dt}\int \rho_t \, U_t \dx =  \int (\partial_t \rho_t) \, U_t \dx + \int \rho_t \, (\partial_t U_t) \dx$ \cite{first-law-of-thermodynamics}, algebraic manipulation yields
\begin{align}
    \frac{\dee W}{\dt} &= -  \int \frac{\partial \rho_t}{\partial t} \, U_t \dx  + \frac{\dee}{\dt} \int \rho_t \, U_t \dx \nonumber \\ 
    &= -  \int \frac{\partial \rho_t}{\partial t} \, (U_t + \ln \rho_t + 1) \dx + \frac{\dee}{\dt} \int \rho_t \, (U_t + \ln \rho_t) \dx  \nonumber \\ 
    &= \int \rho_t | \nabla (U_t + \ln \rho_t)|^2 \dx +  \frac{\dee}{\dt} \int \rho_t \, (U_t + \ln \rho_t) \dx .
\end{align}
Here in the second line we have added and subtracted $\frac{\dee}{\dt} \int \rho_t \ln \rho_t \dx = \int (\partial_t \rho_t ) (\ln \rho_t + 1) \dx$, and in the third line we have plugged in the Fokker-Planck equation Eq.~\eqref{eq:FP-continuity} and integrated by parts in $x$.

Finally, the work is defined as the time-integral of the work rate 
\begin{equation}
  W = \int_0^\tau \frac{\dee W}{\dt} \dt = \int_0^\tau \int \rho_t |\nabla \phi_t|^2 \dx \dt + \int \rho_t (U_t + \ln \rho_t) \dx \bigg|_{t = 0}^\tau, \label{eq:work}
\end{equation}
where we have used the definition of $\phi_t$ in Eq.~\eqref{eq:FP-continuity}.

After noting that the KL divergence from a distribution $\rho$ to another an equilibrium distribution $\rho^\mathrm{eq}_\lambda$ can be equivalently written as
\begin{align}
  D_\mathrm{KL}(\rho | \rho^\mathrm{eq}_\lambda) &:= \int \rho(x) \, \ln \bigg[ \frac{\rho(x) }{\rho^\mathrm{eq}_\lambda (x) } \bigg] \dx = \int \rho \, (\ln \rho + U_\lambda) \dx - F(\lambda)
\end{align}
(recall $F(\lambda) := -\ln \int \exp(-U_\lambda(x)) \dx $ is the equilibrium free energy), the excess work $W_\mathrm{ex} = W - \Delta F$ can be written as
\begin{equation}
  W_\mathrm{ex} = \int_0^\tau \int \rho_t |\nabla \phi_t|^2 \dx \dt + D_{\mathrm{KL}}( \rho_\tau | \rho_f^\mathrm{eq} ), \label{eq:excess-work}
\end{equation}
where we have applied the boundary conditions $\rho_0 = \rho_i^\mathrm{eq}, U_0 = U_i$, and $U_\tau = U_f$. 

The first term in Eq.~\eqref{eq:excess-work} represents the dissipation occurring within the protocol from $t = 0$ and $t = \tau$ \cite{ito2024geometric} and is optimized by a Benamou-Brenier solution between $\rho_0$ and $\rho_\tau$, while the second term represents the dissipation occurring for $t > \tau$ (i.e., after the protocol) as the distribution relaxes from $\rho_\tau$ to $\rho_\infty = \rho^\mathrm{eq}_f$. The change of variables $s = t / \tau$ thus yields $W_\mathrm{ex}^* = \min_{\rho_f} \mathcal{W}_2^2 [ \rho_i^\mathrm{eq}, \rho_f] / \tau + D_\mathrm{KL}(\rho_f | \rho_f^\mathrm{eq})$, which was noted in \cite{chen2019stochastic} to be equivalent to the JKO scheme used to study convergence properties of the Fokker-Planck equation \cite{jordan1998variational} with effective time-step $h = \Delta t / 2$.

If the additional terminal condition $\rho_\tau = \rho_f^\mathrm{eq}$ is additionally imposed, then the second KL divergence term goes away, reproducing the Benamou-Brenier cost function multiplied by $1/\tau$.

\section{Numerical study of linearly-biased double well optimal protocols}

In this section, we provide details of our numerical study of the linearly-biased double well 
\begin{equation}
    U_\lambda(x) = E_0[(x^2 - 1)^2 / 4 - \lambda x],  \label{eq:double-well-potential}
\end{equation}
for $E_0 = 16, \lambda_i = -1, \lambda_f = 1$, and analyze the reduction in performance around $\tau \sim 2$ seen in Fig~S1(a) (a higher resolution version of Fig.~2(e)). We propose an explanation for the geodesic-counterdiabatic protocols overshooting the optimal protocols (Fig~S2(b)) that correspond to a reduction in performance for $\tau \sim 2$.

\subsection{Lattice discretization implementation}

In order to calculate $g(\lambda)$, $h(\lambda),  \mathcal{T}^2(\lambda_i, \lambda), D_\mathrm{KL}(\rho_\lambda^\mathrm{eq} | \rho_f^\mathrm{eq})$, as well as to measure the performance $W_\mathrm{ex}[\lambda(t)]$, we base our numerical implementation on the lattice-discretized Fokker-Planck equation method introduced in \cite{holubec2019physically}. This is the same discretization scheme used by \cite{zhong2022limited}, which allows a direct comparison of numerical results.

We discretize the one-dimensional configuration space as an $N$-state lattice with spacing $\Delta x$ and reflecting boundaries at $x_\mathrm{b} = \pm (N - 1) \Delta x /2$. Following \cite{zhong2022limited}, we use $\Delta x = 0.025$ and $x_b = \pm 3$. The probability density may be represented as a vector $\boldrho(t) = (\rho^1(t), \rho^2(t), ..., \rho^N(t))$ via $\rho(x, t) = (\Delta x)^{-1} [\boldrho(t)]^{l(x)}$ , where $l(x) = \lfloor x / \Delta x + N/2 \rfloor$. Likewise, the potential energy may be represented as a covector $\boldU_{\lambda} = (U_1(\lambda), U_2(\lambda), ... , U_N(\lambda)))^T$ with $U_\lambda(x) = [\boldU_\lambda]_{l(x)}$, yielding the equilibrium probability vector $\boldrho^\mathrm{eq}_\lambda = Z(\lambda)^{-1} (\exp(-U_1(\lambda)), \exp(-U_2(\lambda)), ..., \exp(-U_N(\lambda)))$, where $Z(\lambda) = \sum_i \exp(-U_i(\lambda))$ is the normalization constant. The excess conjugate force is given by the covector 
\begin{equation}
    \boldsymbol{\delta f}(\lambda) = - \frac{\partial \boldU_{\lambda}}{\partial \lambda}  +  \frac{\partial \boldU_{\lambda}}{\partial \lambda}  \cdot \boldrho^\mathrm{eq}_\lambda,
\end{equation}
where the second term is a dot product with the equilibrium probability vector. 

The Fokker-Planck equation is represented as the master equation
\begin{equation}
    \dot{{\boldrho}} = \curlyL_{\lambda} {\boldrho}, \label{eq:master-equation}
\end{equation}
where $\curlyL_\lambda$ is a transition rate matrix on which we impose the following form
\begin{equation}
    {[\curlyL_\lambda]^i}_{j} = 
    \begin{cases} 
    (\Delta x)^{-2} \ e^{ (U_j(\lambda) - U_i(\lambda))/2 }
    & {| i - j| = 1} \\
    - (\Delta x)^{-2} \ \sum_{k \neq j}\  e^{(U_j(\lambda) - U_k(\lambda))/2 }
    & {i = j} \\
    0 & \mathrm{else}.
  \end{cases} \label{eq:transition-rate-matrix}
\end{equation}
Taking the continuum limit $N \rightarrow \infty, \Delta x \rightarrow 0$ with constant $x_\mathrm{b}$ yields the Fokker-Planck equation $\partial_t \rho = \mathcal{L}_\lambda \rho$ \cite{zhong2022limited}.

For this lattice discretization, The friction tensor and Fisher information metric are given by
\begin{equation}
    g(\lambda) = -\sum_{i = 1}^N [\boldrho^\mathrm{eq}_\lambda]^{i} [\boldsymbol{\delta f}]_i \big[ (\curlyL_{\lambda}^T)^{-1} \boldsymbol{\delta f} \big]_i \quad\mathrm{and}\quad h(\lambda) = \sum_{i = 1}^N [\boldrho^\mathrm{eq}_\lambda]^{i} [\boldsymbol{\delta f}]_i [\boldsymbol{\delta f} ]_i, \label{eq:tensors}
\end{equation}
where $(\curlyL_{\lambda}^T)^{-1}$ is the inverse of the adjoint Fokker-Planck matrix with its zero-mode removed \cite{wadia2022solution}. The KL-divergence is 
\begin{equation}
    D_\mathrm{KL}(\rho_\lambda^\mathrm{eq} | \rho_f^\mathrm{eq} ) = \sum_{i = 1}^N [\boldrho^\mathrm{eq}_\lambda]^{i} \, \ln \bigg[ \frac{[\boldrho^\mathrm{eq}_\lambda]^{i}}{[\boldrho^\mathrm{eq}_f]^{i}} \bigg], \label{eq:D_KL}
\end{equation}
and the squared thermodynamic distance (cf., Eq.~(5) in \cite{crooks2007measuring}) takes the form
\begin{equation}
    \mathcal{T}^2(\lambda_i, \lambda) = \bigg[ \int_{\lambda_i}^\lambda \sqrt{g(\lambda)} \, \dee \lambda \bigg]^2. \label{eq:T2}
\end{equation}
We numerically compute the integral as a trapezoid sum on interval $[\lambda_i, \lambda]$ split into 1000 even subintervals.

Geodesic-counterdiabatic protocols of duration $\tau$ are obtained by first solving for $\gamma_f$ through minimizing
\begin{equation}
  \gamma_f = \arg \min_{\lambda} \ \mathcal{T}^2(\lambda_i, \lambda) / \tau + D_\mathrm{KL}(\rho_\lambda^\mathrm{eq} | \rho_f^\mathrm{eq} ) \label{eq:gamma_f_SM}
\end{equation}
using Eq.~\eqref{eq:D_KL} and \eqref{eq:T2} above, and then computing the time-discretized geodesic for $k = 0, 1, 2, ..., K$
\begin{equation}
    \gamma(\cdot) = \{ (s_0 = 0, \gamma_0 = \lambda_i), (s_1, \gamma_1), (s_2, \gamma_2), ..., (s_{K - 1}, \gamma_{K - 1}), (s_K = 1, \gamma_K = \gamma_f) \} 
\end{equation}
using equally spaced $\gamma_k = [1 - (k / K)] \lambda_i + (k / K) \gamma_f = \lambda_i + k \Delta \gamma$, and variable timesteps $s_{k+1} - s_k = \alpha \Delta \gamma \sqrt{g(\gamma_{k}) + g(\gamma_{k + 1})/2}$ with scaling factor $\alpha$ chosen so that $s_K = 1$. Finally, the time-discretized $\eta(\cdot)$ is obtained via 
\begin{equation}
    \eta(s) = h^{-1}(\gamma(s)) g(\gamma(s)) \frac{\dee \gamma(s)}{\ds} \label{eq:cd-term-SM}
\end{equation}
where $\dee \gamma(s) / \ds$ is computed using finite differences. 

Following \cite{zhong2022limited}, we use $K = 1000$. This way of computing geodesic-counterdiabatic protocols is consistent with how geodesic protocols were obtained in \cite{zhong2022limited}, allowing for a direct comparison of performance. 

\subsection{Phase transition for $\gamma_f$}

An interesting property we found for this problem is that the geodesic endpoint $\gamma_f$ obtained as the argmin of the objective function Eq.~\eqref{eq:gamma_f_SM} has a discontinuity at $\tau \approx 3.3$ (Fig.~S2(a)). This is due to the fact that the objective function, computed using Eqs.~\eqref{eq:tensors}-\eqref{eq:T2}, is not convex, and has multiple local minima for $\tau \sim 3$ (Fig.~S2(b)). 

\subsection{Measuring performance}

For a time-discretized protocol $\lambda(t) = \{(t_0 = 0, \lambda_0 = \lambda_i), (t_1, \lambda_1), ..., (t_{K - 1}, \lambda_{K - 1}), (t_K = \tau, \lambda_K)\}$, the trajectory for $\boldrho(t)$ is obtained by integrating Eq.~\eqref{eq:master-equation}
\begin{equation}
    \boldrho_{k+1} = \exp [ \mathcal{L}_{k + 1/2} (t_{k+1} - t_k) ] \, \boldrho_k \quad\mathrm{with}\quad \boldrho_0 = \boldrho_i^\mathrm{eq},
\end{equation}
where $\mathcal{L}_{k + 1/2}$ is the transition rate matrix Eq.~\eqref{eq:transition-rate-matrix} for $\lambda_{k+1/2} = (\lambda_{k} + \lambda_{k+1} / 2)$. 

Finally, the work for a particular protocol is calculated through the first law of thermodynamics
\begin{equation}
    W[\lambda(t)] = \Delta E - Q[\lambda(t)] = [ \boldU_f \cdot \boldrho_K - \boldU_i \cdot \boldrho_i^\mathrm{eq}] - \sum_{k = 0}^{K - 1} \boldU_{k + 1/2} \cdot [ \boldrho_{k+1} - \boldrho_k],
\end{equation}
and the excess work is $W_\mathrm{ex} = W[\lambda(t)] - \Delta F$. Because we are considering $\lambda_i = -1$ and $\lambda_f = 1$ for Eq.~\eqref{eq:double-well-potential}, we have $\Delta F = 0$.

\subsection{Reduction in performance around $\tau \sim 2$}

Though they still outperform geodesic protocols, geodesic-counterdiabatic protocols obtained via Alg.~1 exhibit a noticeable decrease in performance compared with the exact optimal limited-control solutions computed in~\cite{zhong2022limited} using optimal control theory on PDEs. It is important to note the assumptions made for Alg.~1: 
\begin{enumerate}
  \item The complete $L^2$ optimal transport solution, a Wasserstein geodesic $\rho^*_s|_{s \in [0,1]} \in \mathcal{P}(\mathbb{R}^d)$, is closely approximated by a trajectory of equilibrium distributions $\rho^\mathrm{eq}_{\lambda(s)}|_{s \in [0,1]} \in \mathcal{P}^\mathrm{eq}_\mathcal{M}(\mathbb{R}^d)$, i.e., a thermodynamic geometry geodesic.
  \item In the case of limited expressivity of available controls, the continuity equation $\partial_s \rho^\mathrm{eq}_{\gamma(s)} = \eta^\mu(s) \, \nabla \cdot (\rho^\mathrm{eq}_{\gamma(s)} \nabla U_\mu )$ is approximately satisfied by the counterdiabatic driving term obtained via Eq.~\eqref{eq:cd-term-SM}, so that the time-dependent probability distribution $\rho(\cdot, t) \in \mathcal{P}(\mathbb{R}^d)$ solving the Fokker-Planck equation (Eq.~\eqref{eq:FP-continuity}) under the geodesic-counterdiabatic protocol $\lambda^*(t) = \gamma(t / \tau) + \tau^{-1} \eta(t / \tau)$ closely approximates the geodesic path of equilibrium distributions $\rho^\mathrm{eq}_{\gamma(t / \tau)} \in \mathcal{P}^\mathrm{eq}_\mathcal{M}(\mathbb{R}^d)$.
\end{enumerate}
Both of these assumptions hold for the parametric harmonic oscillator, but do not for the linearly-biased double well (Eq.~\eqref{eq:double-well-potential}, $E_0 = 16$) for $\tau \sim 2$. In the case that one or both of these assumptions are broken, Alg.~1 may produce a protocol that poorly approximates the true optimal protocol $\lambda^*(t)$. In the case that the first assumption is broken, the global optimal limited-control protocol $\lambda^*(t)$ (i.e., obtained using optimal control on PDEs \cite{zhong2022limited}) may yield a Fokker-Planck equation solution $\rho^*(\cdot, t)|_{t \in [0, \tau]}$ that better approximates the optimal transport solution $\rho^*_s|_{s \in [0, 1]}$ by not being on the equilibrium manifold, i.e., $\rho^*(\cdot, t)|_{t \in [0, \tau]} \notin \mathcal{P}(\mathbb{R}^d)$.

We believe that this existence of multiple local minima for $\gamma_f$ (Fig.~S2(b)) indicates that near values of this ``critical protocol duration'' $\tau \approx 3.3$, the first assumption is indeed broken. The objective function yielding the terminal time $\rho^*_1 = \rho_f^*$ for the optimal transport solution 
\begin{equation}
  \rho_f^* = \arg \min_{\rho_f} \, \mathcal{W}_2^2[\rho_i, {\rho_f}] / \tau + D_{\mathrm{KL}}( {\rho_f} | \rho_f^\mathrm{eq} )
\end{equation}
has only a single minimum, as constructed and proved in \cite{jordan1998variational}, and therefore is unlikely to represent an equilibrium distribution for any choice of control parameter values $\rho_f^* \notin \mathcal{P}_\mathcal{M}^\mathrm{eq}(\mathbb{R}^d)$. 
Thus, for $\tau \sim 2$, the optimal transport solution $\rho^*_s|_{s \in [0, 1]}$ is not well approximated by an equilibrium distribution trajectory, and so Alg.~1 does not produce a protocol that approximates the optimal protocol $\lambda^*(t)$ obtained in \cite{zhong2022limited} (Fig.~S1(b)), leading to a decrease in performance (Fig.~S1(a)). Nevertheless optimal protocols obtained via Alg.~1 still noticeably outperform the geodesic protocol that connect $\lambda_i$ to $\lambda_f$ (Fig.~S1(a)).

\begin{figure}
\includegraphics[width=14cm]{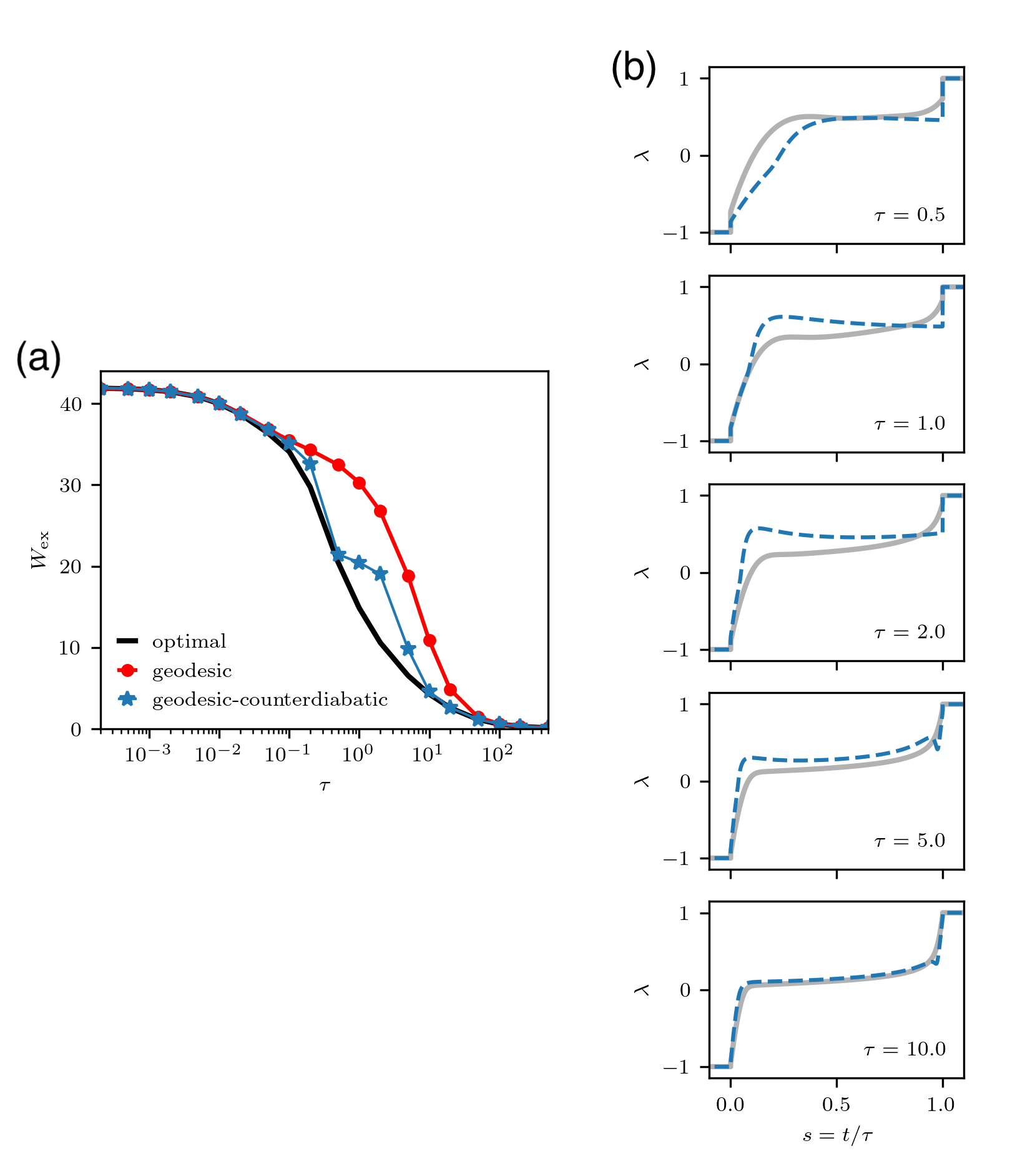}
\caption{(a) Performance of optimal protocols from \cite{zhong2022limited} (black), geodesic (red dots), and geodesic-counterdiabatic protocols (blue stars), higher resolution reproduction of Fig.2(c) from main text.  (b) Optimal protocols from \cite{zhong2022limited} (filled grey line) and geodesic-counterdiabatic protocols from Alg.~1 (broken blue line) for $\tau$ between $0.5$ to $10.0$, corresponding to the decrease in performance ``peak'' in (a). We see that for $\tau \sim 2$, the geodesic-counterdiabatic protocols overshoot the optimal protocols, corresponding to a reduction in performance.}  \label{fig:SI-1}
\end{figure}

\begin{figure}
\includegraphics[width=14cm]{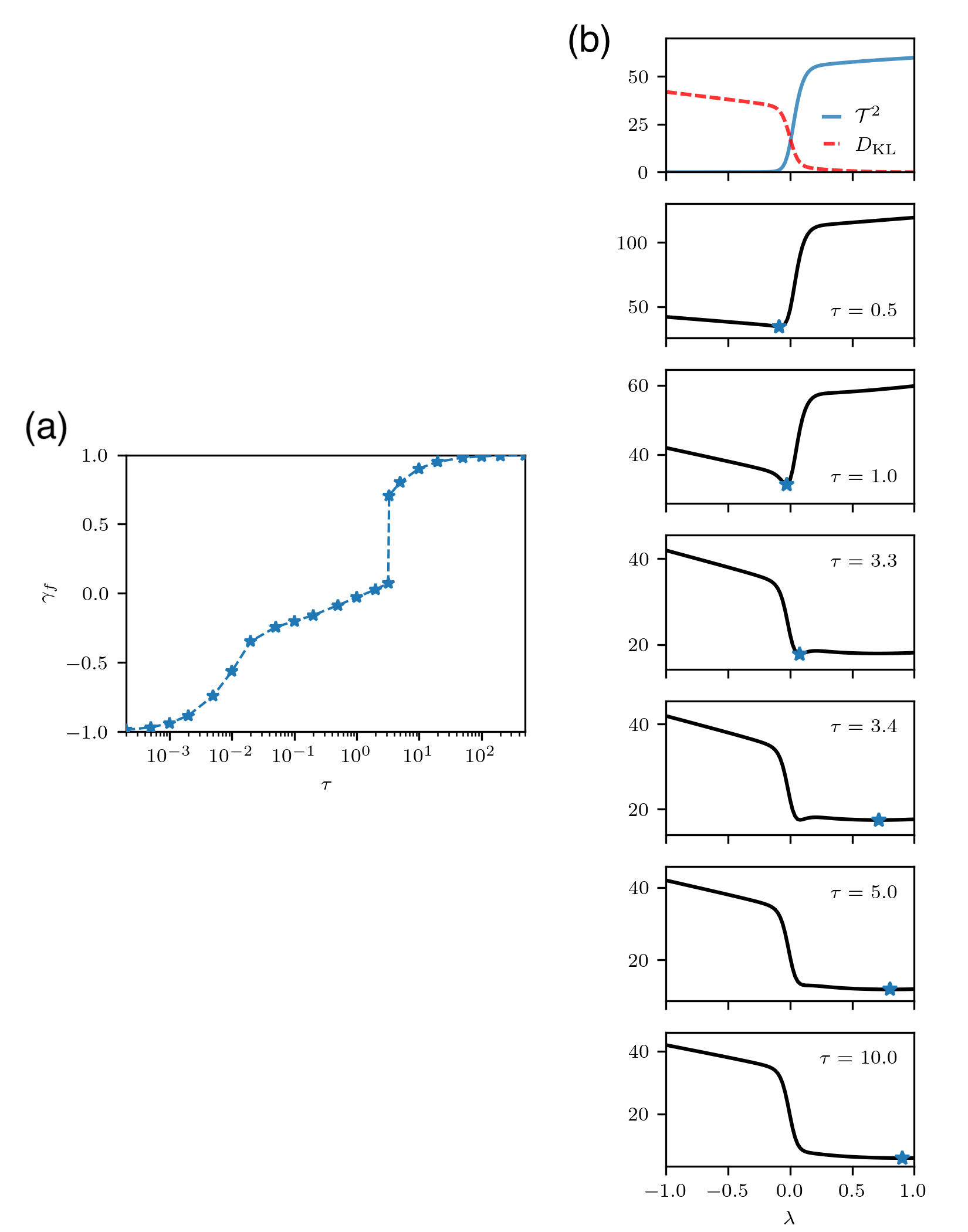}
\caption{(a) The geodesic endpoint for geodesic-counterdiabatic protocols $\gamma_f = \arg \min_{\lambda} \ \mathcal{T}^2(\lambda_i, \lambda) / \tau + D_\mathrm{KL}(\rho_\lambda^\mathrm{eq} | \rho_f^\mathrm{eq} ) $, as a function of protocol duration $\tau$. At $\tau \approx 3.3$ there is a discontinuity from $\gamma_f = 0.073$ to $\gamma_f = 0.706$. (b) The functions $\mathcal{T}^2(\lambda_i, \lambda)$ (solid blue) and $D_\mathrm{KL}(\rho_\lambda^\mathrm{eq} | \rho_f^\mathrm{eq} )$ (broken red), and their sum $\mathcal{T}^2(\lambda_i, \lambda) / \tau + D_\mathrm{KL}(\rho_\lambda^\mathrm{eq} | \rho_f^\mathrm{eq} )$ for various values of $\tau$. The blue star indicates the minimum, yielding $\gamma_f$, which suddenly jumps from $\gamma_f = 0.073$ at $\tau = 3.3$ to $\gamma_f = 0.706$ at $\tau = 3.4$, due to the existence of multiple local minima in the objective function.}  \label{fig:SI-2}
\end{figure}

\bibliography{main}